\documentclass[twocolumn,english,aps,prl,showpacs,superscriptaddress]{revtex4-1}
\usepackage[LGR,T1]{fontenc}
\usepackage[latin9]{inputenc}
\usepackage{units}
\usepackage{amsmath}
\usepackage{amssymb}
\usepackage{graphicx}
\usepackage{wasysym}

\makeatletter

\newcommand*\LyXThinSpace{\,\hspace{0pt}}
\DeclareRobustCommand{\greektext}{%
  \fontencoding{LGR}\selectfont\def\encodingdefault{LGR}}
\DeclareRobustCommand{\textgreek}[1]{\leavevmode{\greektext #1}}
\ProvideTextCommand{\~}{LGR}[1]{\char126#1}

\usepackage{babel}
\usepackage{dsfont}

\usepackage{hyperref}
\def\equationautorefname~#1\null{Equation (#1)\null}

\makeatother

\usepackage{babel}
\begin{document}

\title{Transverse optical pumping of spin states}

\author{Or Katz}
\email[Corresponding author:]{ or.katz@weizmann.ac.il }

\affiliation{Department of Physics of Complex Systems, Weizmann Institute of Science,
Rehovot 76100, Israel}

\affiliation{Rafael Ltd, IL-31021 Haifa, Israel}

\author{Ofer Firstenberg}

\affiliation{Department of Physics of Complex Systems, Weizmann Institute of Science,
Rehovot 76100, Israel}
\begin{abstract}
Optical pumping is an efficient method for initializing and maintaining
atomic spin ensembles in a well-defined quantum spin state. Standard
optical-pumping methods orient the spins by transferring photonic
angular momentum to spin polarization. Generally the spins are oriented
along the propagation direction of the light due to selection rules
of the dipole interaction. Here we present and experimentally demonstrate
that by modulating the light polarization, angular momentum perpendicular
to the optical axis can be transferred efficiently to cesium vapor.
The transverse pumping scheme employs transversely oriented dark states,
allowing for control of the trajectory of the spins on the Bloch sphere.
This new mechanism is suitable and potentially beneficial for diverse
applications, particularly in quantum metrology.
\end{abstract}
\maketitle
Optical pumping is the prevailing technique for orienting atomic spins,
conveying order from polarized light onto the state of spins \citep{Happer-1972,Happer-book,Auzinsh-book}.
Many applications in precision metrology \citep{Budker-Romalis-optical-magnetometery,Serf-magnetometer,Kitching-sensors,Treutlein-microwave-imaging},
quantum information \citep{Polzik-RMP-2010,POlzik2017,Novikova-review},
noble gas hyper-polarization \citep{happer-walker-rmp,happer-1998,Chupp},
and searches for new physics beyond the standard model \citep{fundamental-physics,romalis-fundamental-physics}
employ optical pumping for initializing the orientation moment of
the spins, that is, for pointing the spins towards a preferred direction.
The required degree of polarization depends on the specific application,
where optimized performance in quantum metrology is often practically
achieved around 50\% polarization \citep{romalis-hybrid,fifty_precent_polarization,Muschik}.
Standard optical pumping schemes generate polarization along the propagation
direction of the laser beam. These schemes include depopulation pumping
\citep{Happer-1972}, synchronous pumping \citep{Weis-PM-1,Romalis-2007,SYNCHRONEOUS-PUMPING2},
spin-exchange indirect pumping \citep{Chalupczak-indirect-SE-pumping,Chalupczak-spin-maser},
alignment-to-orientation conversion \citep{Budker-AOC}, and hybrid
spin-exchange pumping \citep{romalis-hybrid}. However in various
applications, it is often desired to polarize the spins along an applied
magnetic field, perpendicular to the optical axis \citep{Muschik,Budker-book,kitching-pump-probe,Walker-NMRG,SERF-storage-of-light}.
While at extreme magnetic fields, it is possible to polarize the spins
transversely \citep{footnote1}, at moderate magnetic fields, typical
to alkali-metal spins experiments for example, the pumping efficiency
is rather low. 

Here we propose and demonstrate an optical pumping scheme for efficient
spin polarization transversely to the propagation direction of the
laser beam. The scheme incorporates a polarization-modulated light
beam, which steers the spins in helical-like trajectories on the Bloch
sphere around and along a transverse magnetic field, while gradually
increasing their polarization. The scheme exhibits sharp resonances,
reaching maximum efficiency when the optical modulation is resonant
with the Larmor precession of the spins. We develop a simple analytical
model for analyzing the experimental results and discuss the applicability
of the scheme for various applications.

\begin{figure}[b]
\begin{centering}
\includegraphics[viewport=10bp 85bp 820bp 535bp,clip,width=8.5cm]{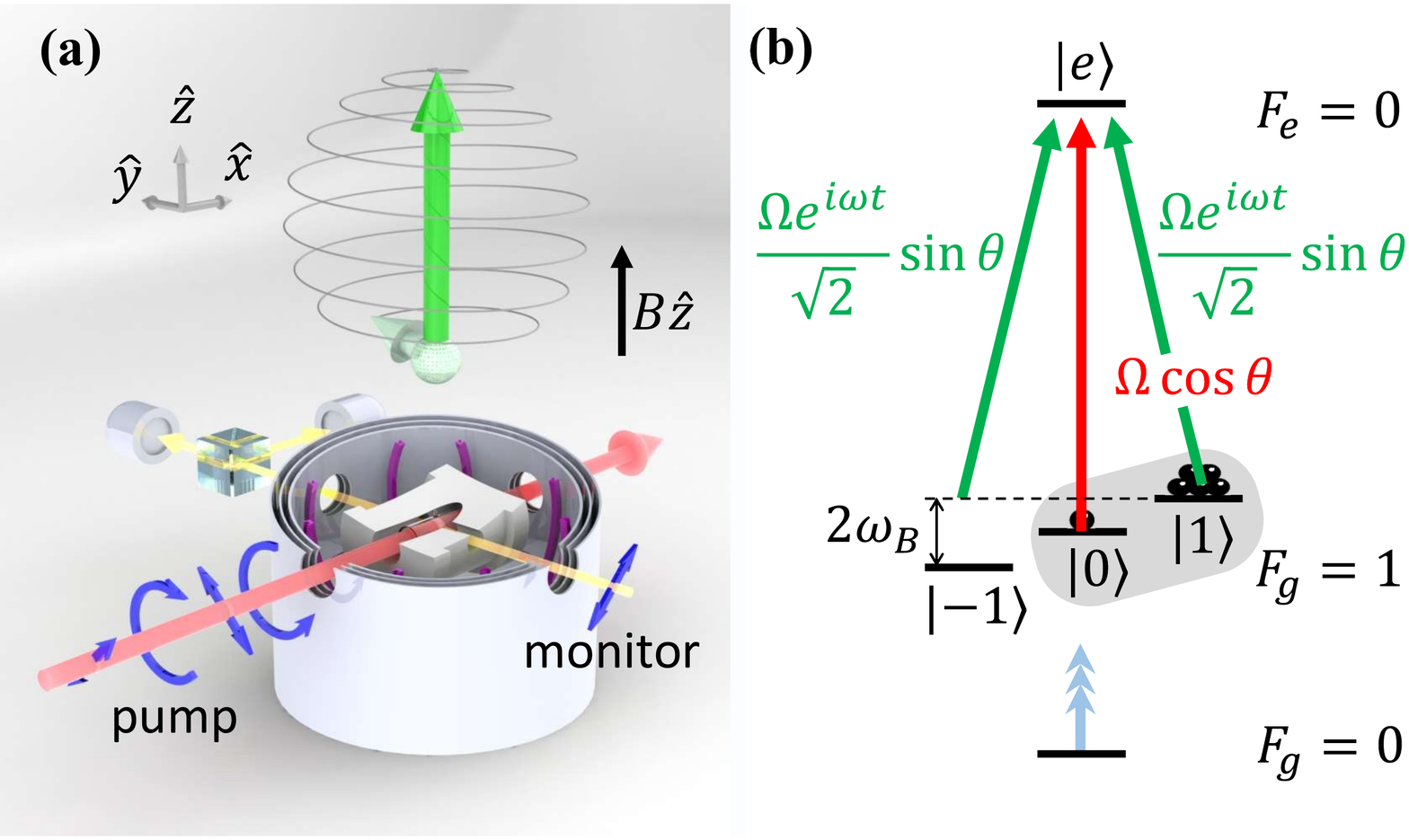}
\par\end{centering}
\centering{}\caption{\textbf{Experimental system and toy model.} (a) Schematics of the
experimental setup and the spiral motion of the atomic spins (green)
towards $+\hat{z}$. The polarization of the pump beam (red) alternates
between linear and circular (blue arrows). The spin orientation is
monitored using balanced polarimetry of a far-detuned monitor beam
(yellow). Not shown are the repump beam and a second monitor beam,
which co-propagate with the pump. (b) Toy model with $I=1/2$. The
repump laser (blue arrow) empties the $F_{g}=0$ state, while the
pump laser (red and green arrows) drives the atoms into the dark state
$|d_{+}\rangle\propto\cos\theta\left|1\right\rangle -\tfrac{1}{\sqrt{2}}e^{i\omega t}\sin\theta\left|0\right\rangle $
(gray shading), eventually oriented perpendicular to the beam direction.\label{fig: exp_system}}
\end{figure}

In standard optical pumping schemes, the atomic ground state is polarized
via repeated cycles of absorption and spontaneous emission. Ideally,
the atoms cease to absorb the pump photons when they reach a `dark
state', which is determined by the excited transitions during pumping
\citep{Happer-1972}. For a light field with an electric field $\mathbf{E}\left(t\right)=E_{0}e^{i(\omega_{\mathrm{L}}t-kx)}\hat{e}$,
the relevant transitions depend on the relative detuning of the light
frequency $\omega_{\mathrm{L}}$ from the atomic transition frequency
$\omega_{0}$, on the external electric and magnetic fields, and on
the selection rules of the dipole interaction for polarization $\hat{e}$.
For alkali-metal vapors, the latter dominates the pumping process
of spin orientation at moderate magnetic fields, because the ground
and excited magnetic sublevels $|F_{g},m_{g}\rangle,\,|F_{e},m_{e}\rangle$
within each hyperfine manifold $F_{g},\,F_{e}$ are optically unresolved. 

For constant polarization $\hat{e}$, one-photon absorption of light
does not produce a considerable spin orientation transversely to the
optical axis. Circular light polarization $\hat{e}_{\pm}=(\hat{y}\pm i\hat{z})/\sqrt{2}$
orients the spins along the optical axis $\pm\hat{x}$ via the allowed
transitions $m_{e}=m_{g}\pm1$; For $F_{e}\le F_{g}$, the maximally
polarized state $|m_{g}=\pm F_{g}\rangle$ is dark. Linearly polarized
light $\hat{e}=\hat{y},\hat{z}$ generates spin alignment along $\hat{x}\times\hat{e}$
and zero net orientation with the selection rules $m_{e}=m_{g}$ when
tuned to the transition $F_{g}\rightarrow F_{e}=F_{g}-1$; This generates
a quadrupole magnetic moment \citep{Happer-1972}, leaving both $|m_{g}=F_{g}\rangle$
and $|m_{g}=-F_{g}\rangle$ dark. It thus seems that no orientation
is built perpendicularly to the optical axis $\hat{x}$ for any light
polarization. Our scheme overcomes this limitation and allows for
transverse optical pumping of the spins by temporally modulating the
light polarization. 
\begin{figure}[t]
\begin{centering}
\includegraphics[viewport=0bp 0bp 390bp 372bp,clip,width=8.6cm]{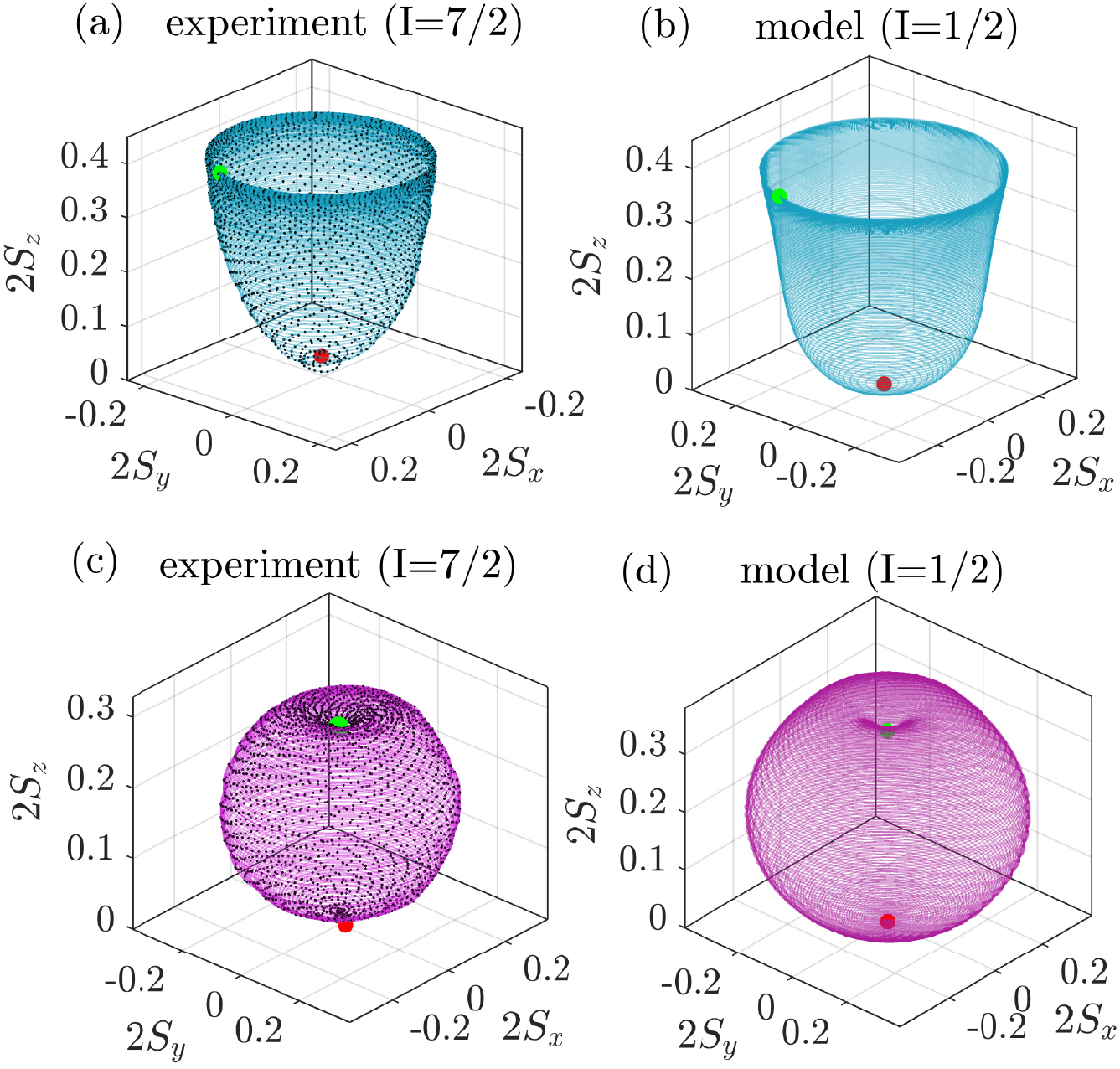}
\par\end{centering}
\centering{}\caption{\textbf{Spin trajectories on the Bloch sphere. }The optical axis is
$\hat{x}$, within the equatorial plane. \textbf{(a,c)} Measurements
of the pumping process from $t=0$ to $t=100$ ms. \textbf{(b,d)}
Theoretical toy model with $I=1/2$. Red (green) circles mark the
initial (final) states of the spin. \textbf{(a)} When pumping with
a constant modulation depth $\theta=0.2\;\mbox{rad}$, the measured
cesium spins follow a spiral-helical trajectory around the $+\hat{z}$
direction. \textbf{(c)} Adiabatically varying $\theta\left(t\right)$
allows for driving the spins in a spherical-helical trajectory that
ends along the $\hat{z}$ axis.\label{fig: Helix} }
\end{figure}

\section*{Results}

We employ the experimental setup shown schematically in Fig.~\ref{fig: exp_system}(a),
containing cesium vapor at room temperature. Setting a constant magnetic
field $B\hat{z}$ determines the quantization axis $\hat{z}$ and
the Larmor frequency $\omega_{\mathrm{B}}=gB$, where $g=0.35\,(2\pi)\text{MHz/G}$
is the gyro-magnetic ratio for cesium. For the transverse pumping,
we use a\textit{ pump} beam, which frequency is tuned to the $D_{1}$
transition $F_{g}=4\rightarrow F_{e}=3$ and which polarization is
modulated according to 
\begin{equation}
\hat{e}\left(t\right)=\cos\left(\theta\right)\hat{z}+ie^{i\omega t}\sin\left(\theta\right)\hat{y}.\label{eq: Beam_modulation}
\end{equation}
Here $\sin\left(\theta\right)$ is the modulation depth and $\omega$
is the modulation angular frequency. For the sake of analysis and
presentation, we introduce two far-detuned \textit{monitor} beams
propagating along $\hat{x}$ and $\hat{y}$, measuring the three dimensional
orientation state of the spins $(2S_{x},2S_{y},2S_{z})$ on the Bloch
sphere during the pumping process. See Methods for additional experimental
details.

\begin{figure}[t]
\centering{}\includegraphics[viewport=0bp 0bp 431bp 358bp,clip,width=8.6cm]{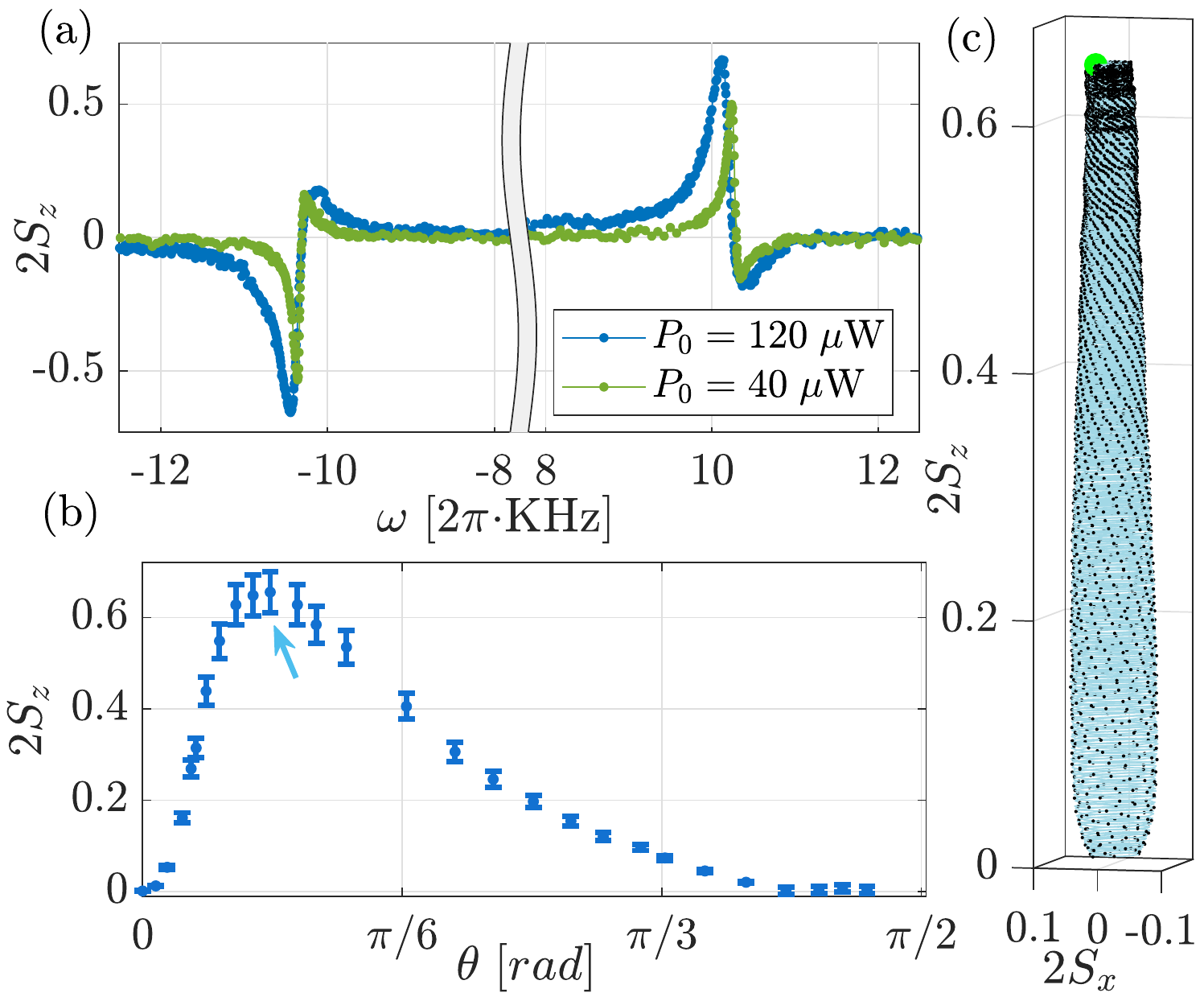}
\caption{\textbf{Pumping dependence on the modulation parameters. (a)} Measured
$S_{z}\left(\omega\right)$ at $t=200\;\mbox{msec}$ for cesium atoms
with $\theta=0.24\;\mbox{rad}$ and $\omega_{\mathrm{B}}=10.2\;\mbox{(2\ensuremath{\pi})kHz}$.
The resonance peaks at $\omega\approx\pm\omega_{\mathrm{B}}$ are
associated with the two CPT dark states $|d_{\pm}\rangle$. \textbf{(b)}
Dependence on the modulation depth $\theta$: measured $S_{z}\left(\theta\right)$
at $t=200\;\mbox{msec}$ with $\omega=\pm10.3\;(2\pi)\text{kHz}$
and $P_{0}=120\;\mbox{\ensuremath{\mu}W}$. Pumping is optimal at
moderate modulation depths, such that $|d_{+}\rangle$ is oriented
towards $+\hat{z}$, but $\Gamma\sin^{2}\left(\theta\right)\apprge\gamma$.
\textbf{(c)} Measured spin trajectory, corresponding to the point
point marked by an arrow in \textbf{b}. The final polarization (green
circle) is along $\hat{z}$, \emph{i.e. }perpendicular to the optical
axis, with a small residual polarization along $\hat{x}$ and $\hat{y}$.\label{fig: Modulation_parameters}}
\end{figure}

A typical measurement during the pumping process is presented on the
Bloch sphere in Fig.~\ref{fig: Helix}(a). We observe that the spin
orientation follows a helical trajectory transversely to the optical
axis $\hat{x}$. In this experiment, the pump power is $P_{0}=250\;\mbox{\ensuremath{\mu}W}$
and the modulation frequency is tuned to resonate with the Larmor
frequency $\omega_{\mathrm{B}}\approx\omega=1.5\;(2\pi)\mbox{kHz}$.
The final value of $2S_{z}$ quantifies the pumping efficiency. Its
dependence on the modulation parameters $\omega$ and $\theta$ is
shown in Fig.~\ref{fig: Modulation_parameters}. We identify two
resonant features of $2S_{z}\left(\omega\right)$ at $\omega\approx\pm\omega_{\mathrm{B}}$
as shown in Fig.~\ref{fig: Modulation_parameters}(a) for $\theta=0.24$
and two laser powers. The laser power governs the width of the resonance
as well as the shift of the peak from the actual Larmor frequency.
Figure \ref{fig: Modulation_parameters}(b) presents $2S_{z}\left(\theta\right)$
on one of the resonances $[\omega=10.3\;(2\pi)\mbox{KHz]}$. We achieved
an overall maximal polarization of $2S_{z}=65\%$ (with small residual
transverse polarization $2${\small{}$\sqrt{S_{x}^{2}+S_{y}^{2}}$}$=3.5\%$)
as shown in Fig.~\ref{fig: Modulation_parameters}(c).

\begin{figure}[t]
\begin{centering}
\includegraphics[width=8.6cm]{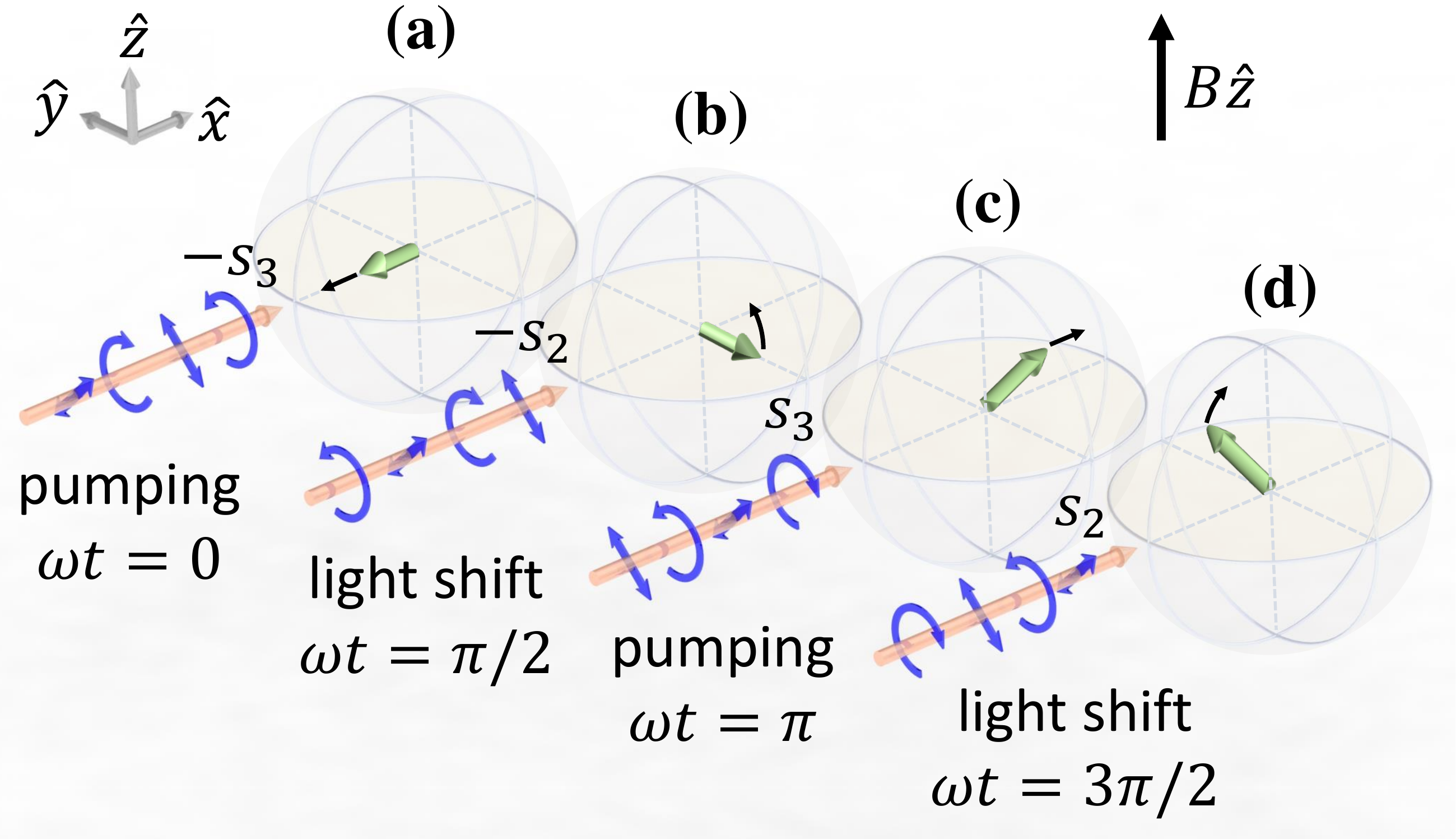}
\par\end{centering}
\centering{}\caption{\textbf{Transverse pumping mechanism.} One period of the transverse
pumping mechanism for synchronous modulation $\omega=\omega_{\mathrm{B}}$.
The spins (green arrows) precess $\omega_{\mathrm{B}}t=\pi/2$ radians
between each subplot due to the magnetic field $B\hat{z}$. (a) The
spins are optically-pumped towards $-\hat{x}$. (b+d) The spins are
tilted towards $+\hat{z}$ due to tensor light-shift. (c) The spins
are optically-pumped towards $+\hat{x}$. Black arrows indicate the
pumping/tilting directions. The eventual spin orientation is along
$+\hat{z}$, \textit{perpendicular} to the propagation direction of
the pumping beam. \label{fig: Scheme_spheres}}
\end{figure}

To explain the transverse pumping mechanism we utilize a simple model
of an alkali-like level structure with nuclear spin $I=1/2$, as shown
in Fig.~\ref{fig: exp_system}(b). The magnetic field $\mathbf{B}=B\hat{z}$
(henceforth assume $B>0$) breaks the isotropy in the transverse $yz$
plane, setting our quantization axis $\hat{z}$ and splitting the
Zeeman sublevels $\left|0\right\rangle ,\left|\pm1\right\rangle $
by $\hbar\omega_{\mathrm{B}}$. The $F_{g}=0$ level is emptied by
a repump field or by spin-exchange collisions. The pump field, resonant
with the $F_{g}=1\rightarrow F_{e}=0$ transition, is polarization-modulated
according to Eq.~(\ref{eq: Beam_modulation}). We describe the effect
of the polarization modulation on the spins dynamics by decomposing
the polarization vector $\hat{e}\left(t\right)$ into its Stokes components
$\boldsymbol{\hat{s}}=(s_{1},s_{2},s_{3})$ \citep{Romalis-Dead-Zone-Free-magnetometer}.
The unmodulated linear polarization $\hat{z}$, represented by $s_{1}$,
aligns the atoms along $\hat{z}$ at a rate $R_{a}\sim\Gamma\cos^{2}\left(\theta\right)$,
creating spin-alignment (see Supplementary Note). Here $\Gamma=\Omega^{2}/\gamma_{\mathrm{e}}$
is the characteristic pumping rate, with $\gamma_{\mathrm{e}}$ the
spontaneous emission rate and $\Omega$ the Rabi frequency of the
pump beam. The linear polarization $(\hat{y}\pm\hat{z})/\sqrt{2}$,
represented by $s_{2}$, induces a tensor light shift of $\Gamma\sin\left(2\theta\right)\sin\left(\omega t\right)$
along $\pm\hat{x}$, which acts like a magnetic field. The circular
polarization $\hat{e}_{\pm}$, represented by $s_{3}$, pumps the
spins longitudinally along $\pm\hat{x}$ at a rate $\Gamma\sin\left(2\theta\right)\cos\left(\omega t\right)$.
Therefore, the modulated polarization alternates between pumping ($s_{3}$)
and light-shifting ($s_{2}$) the atomic spins along $\hat{x}$ at
a rate $\omega$. For $\omega=\omega_{\mathrm{B}}$, the pumping and
light shifts are synchronous, efficiently driving the precessing spins
away from the $xy$ plane, transversely to the optical axis. The resulting
evolution of the Bloch vector $(2S_{x},2S_{y},2S_{z})$ is shown graphically
in Fig.~\ref{fig: Scheme_spheres} and further detailed in Methods.

The toy model enables to reconstruct the main features of the measured
trajectories as shown in Fig.~\ref{fig: Helix}(b), by solving the
$I=1/2$ model numerically and tuning its parameters (see Supplementary
Note). We note however that this model only aims at explaining the
qualitative features of the process, while disregarding effects arising
from the multilevel structure of cesium, which may reduce the pumping
efficiency. 

The resonant nature of the pumping process can also be understood
as originating from coherent population trapping (CPT) \citep{Fleischhauer-EIT-RMP}.
In CPT, a dark state is formed within a $\Lambda$ level-system via
destructive interference of two excitation pathways. Considering the
level structure in Fig.~\ref{fig: exp_system}(b) and decomposing
the modulated pump into its two polarization components $E\hat{z}$
and $E\hat{y}$, we identify two $\Lambda$ systems: $\Lambda_{+}=\{\left|1\right\rangle ,\left|e\right\rangle ,\left|0\right\rangle \}$
and $\Lambda_{-}=\{\left|-1\right\rangle ,\left|e\right\rangle ,\left|0\right\rangle \}$.
System $\Lambda_{+}$ has the dark state $|d_{+}\rangle\propto\cos\left(\theta\right)\left|1\right\rangle -\tfrac{1}{\sqrt{2}}e^{i\omega t}\sin\left(\theta\right)\left|0\right\rangle $
at $\omega\approx\omega_{\mathrm{B}}$, while system $\Lambda_{-}$
has the dark state $|d_{-}\rangle\propto\cos\left(\theta\right)\left|-1\right\rangle -\tfrac{1}{\sqrt{2}}e^{i\omega t}\sin\left(\theta\right)\left|0\right\rangle $
at $\omega\approx-\omega_{\mathrm{B}}$. For $\theta\ll1$, the dark
states $|d_{\pm}\rangle\approx\left|\pm1\right\rangle $ represent
the polarized states perpendicular to the optical axis. The application
of magnetic field $B\hat{z}$ separates the CPT resonances of $\Lambda_{+}$
and $\Lambda_{-}$ by $2\omega_{\mathrm{B}}$, so that the states
$|d_{+}\rangle$ and $|d_{-}\rangle$ cannot be simultaneously dark
when $\omega_{\mathrm{B}}\gg\Gamma$. Consequently, setting $\omega=\omega_{\mathrm{B}}$
depopulates the $\Lambda_{-}$ system while pumping the $\Lambda_{+}$
system towards the transversely oriented dark state $|d_{+}\rangle$.
We conclude that destructive interference of two excitation pathways
effectively modifies the absorption selection-rules, such that one
polarized state (\textit{e.g.} $\left|-1\right\rangle $ for $\omega_{\mathrm{B}}>0$)
absorbs photons, while the opposite state ($\left|1\right\rangle $
for $\omega_{\mathrm{B}}>0$) is transparent. 

We associate the two resonances in Fig.~\ref{fig: Modulation_parameters}(a)
with the CPT dark states of $\Lambda_{+}$ at $\omega>0$ and $\Lambda_{-}$
at $\omega<0$. In the absence of the upper hyperfine level $F_{e}=4$,
the resonances would have seen like single peaks; it is the presence
of $F_{e}=4$ that introduces the second peaks (with opposite sign)
in each resonance, as well as the shift of the peak from $\omega=\omega_{\mathrm{B}}$
and the asymmetry between positive and negative $\omega$. As seen
in Fig.~\ref{fig: Modulation_parameters}(b), the pumping is most
efficient at moderate modulation depths: For $\theta\rightarrow\pi/2$
the dark state is $|d_{+}\rangle\rightarrow\left|0\right\rangle $,
with zero net orientation. For $\theta\rightarrow0$, the dark state
is $|d_{+}\rangle\approx\left|1\right\rangle $, but the depopulating
rate of $|d_{-}\rangle$, proportional to $\Gamma\sin^{2}\left(\theta\right)$,
is too small compared to the overall depolarization rate $\gamma$.

A benefit of the CPT resonant operation is the ability to temporally
vary the system state in a controlled, adiabatic manner. To demonstrate
this, we monitor the pumping process on resonance {[}$\omega=\omega_{\mathrm{B}}=1.5\;(2\pi)\mbox{KHz}${]},
while temporally varying $\theta$ over a duration $T=100\;\mbox{msec}$
according to $\theta\left(t\right)=\arccos\sqrt{t/T}$. The spin state,
initially pumped to $|d_{+}\rangle{}_{\theta\left(t=0\right)}\approx\left|0\right\rangle $,
adiabatically follows the varying dark-state $|d_{+}\rangle{}_{\theta\left(t\right)}$
to its final value $|d_{+}\rangle{}_{\theta\left(t=T\right)}\approx\left|1\right\rangle $,
tracing a spherical-like trajectory as shown in Fig.~\ref{fig: Helix}(c)
(experiment) and Fig.~\ref{fig: Helix}(d) (theory). This process
is similar to stimulated Raman adiabatic passage \citep{stirap} and
can therefore be used to tailor desired trajectories and final states.
Notably it enables the zeroing of the transverse spin components $S_{x}$
and $S_{y}$ at the end of the process, as shown in Figs.~\ref{fig: Helix}(c,d). 

\section*{Discussion}

It is relatively simple to implement the presented scheme in applications.
Polarization modulation can be done using a single photo-elastic modulator
\citep{Hinds} or with readily available, on chip, integrated photonics
\citep{Op5tical-chip}. Various applications that rely on optical
spin manipulation would potentially benefit from utilizing the scheme.
Here we briefly consider some directions with spin vapors. 

First, devices currently employing perpendicular beams in a pump-probe
configuration \citep{Budker-book,kitching-pump-probe,Walker-NMRG}
could be realized with a simpler, co-propagating arrangement, with
the spins oriented transversely to the optical axis. Such arrangement
is most beneficial for miniaturized sensors, such as NMR-oscillators
\citep{Walker-NMRG}, where the size and complexity depend crucially
on the beams configuration, especially if the light source, manipulation,
and detection can be implemented on a single stack over a chip \citep{kitching-pump-probe}.
These sensors are used in various applications as well as in fundamental
research, such as search for new physics \citep{WALKER-pt-ciolation-nmrg,budker-optical-magnetometer}.
Particularly for NMR-oscillators, the projection of the alkali spin
along the magnetic field will be unmodulated, thus sustaining the
spin-exchange optical-pumping of the noble gas spins. 

Second, any application that is restricted to a single laser direction
can now use our scheme to control and fine tune the final direction
of the pumped spins (longitudinal, transversal, or combination thereof).
Such applications include remote magnetometry of mesospheric sodium
spins \citep{Budker-book,budker-mesospheric}, steady-state entanglement
generation during pumping \citep{Muschik}, optical pumping of metastable
$^{3}\text{He}$ for medical imaging \citep{Walker-He3}, and coherent
manipulation of the internal spin state of cold atoms without associate
heating \citep{Footnote3}. 

Third, transverse pumping may form the basis for an all-optical magnetometer
using either alkali-metal atoms or metastable $^{4}\text{He}$ atoms
designed for space applications \citep{Budker-book,budker-optical-magnetometer}.
This magnetometer would rely on measuring the resonant response to
the modulated light, providing a dead-zone-free operation \citep{Romalis-Dead-Zone-Free-magnetometer},
or on measuring the Faraday-rotation of off-resonant probe light,
thus reducing the photon shot-noise commonly limiting magnetometers
based on electromagnetically induced transparency \citep{Fleischhauer-EIT-magnetometer-limit}.
Moreover, the polarization-modulated pump generates the $m=1$ Zeeman
coherence \citep{SERF-storage-of-light}, implying that these magnetometers
could operate in the spin-exchange relaxation-free (SERF) regime \citep{nonlinear-SERF}. 

Finally, our scheme does not rely on any process particular to vapor
physics. It is thus readily application to any spin system having
a non-degenerate \textgreek{L}-system with a meta-stable ground manifold,
such as those employed in diamond color centers \citep{nv_CENTERS,nv-center2},
rare-earth doped crystals \citep{RARE-EARTH}, and semiconductor quantum
dots \citep{QDot,QDot2,SIV-Nunn}. 

In conclusion, we have demonstrated a new optical pumping technique,
generating significant spin orientation transversely to the propagation
direction of the pump beam. The spins are oriented along the external
transverse magnetic field via alternating actions of pumping and tensor
light-shifts, which are resonant with the Larmor precession. The resonance
features, associated with transversely orientated dark-states, allow
one to control the spin trajectory on the Bloch sphere by varying
the modulation parameters. This scheme could be highly suitable for
quantum-metrology applications.

\section*{Methods}

\subsection*{Additional experimental details:}

We use a $10$-mm diameter, $30$-mm long cylindrical glass cell containing
cesium vapor ($I=7/2,\,S=1/2$) at room temperature. The cell is paraffin
coated and free of buffer gas, exhibiting spin coherence time of 150
ms \citep{SERF-storage-of-light}. We set a constant magnetic field
$B\hat{z}$ in the cell using Helmholtz coils and four layers of magnetic
shields. For the transverse pumping, we use an $895\;\mbox{nm}$ single-mode
\textit{pump} beam. We modulate the pump polarization by splitting
it with a polarizing beam splitter (PBS) and sending each output arm
to an acousto-optic modulator. The beams deflected by the modulators
are recombined and mode-matched using a second PBS, resulting with
the polarization given in Eq.~\ref{eq: Beam_modulation}. The laser
frequency after passing the modulators is tuned to the $D_{1}$ transition
$F_{g}=4\rightarrow F_{e}=3$. We control the modulation depth $\sin\left(\theta\right)$
by rotating the linear polarization before the first PBS and control
the modulation angular frequency $\omega$ by setting the relative
RF frequencies of the two modulators. To keep the lower hyperfine
manifold $F_{g}=3$ empty, we use 1 mW of auxiliary \textit{repump}
beam at $895\;\mbox{nm}$, resonant with the $F_{g}=3\rightarrow F_{e}=4$
transition and linearly polarized along $\hat{y}$. The pump and repump,
both with a diameter of 8 mm, co-propagate along $\hat{x}$. 

\subsection*{Reconstruction of the spin state on the Bloch sphere: }

The spin state is reconstructed by evaluating the electronic spin
orientations $(2S_{x},2S_{y},2S_{z})=\left\langle \mathbf{F}\right\rangle /4$,
where $\left\langle \mathbf{F}\right\rangle $ is the orientation
moment of the total spin operator $\mathbf{F}=\mathbf{I+S}$. The
spin orientations are measured by using balanced polarimetry of two
linearly-polarized far-detuned \textit{monitor} beams propagating
along $\hat{x}$ and $\hat{y}$. At low atomic densities and depopulated
$F_{g}=3$ hyperfine manifold, the detected Faraday-rotation angles
are proportional to the spin orientation along the direction of the
beam \citep{Happer-book,coherent-coupling}. We calibrate the proportionality
constants of each monitor beam by measuring its maximal polarization
rotation when the ground state is fully pumped using two circularly-polarized
beams resonant with the two ground-state hyperfine manifolds. We reconstruct
the three spin components by making two consecutive measurements:
First, $S_{x}$ and $S_{y}$ are measured when $\mathbf{B}=B\hat{z}$.
Second, a measurement is conducted with $\mathbf{B}=B\hat{y}$ and
$\theta$ changed by $\theta\rightarrow\pi/2-\theta$, keeping the
other experimental parameters unchanged. As a result, spin is built
along $\hat{y}$ and measured by the $\hat{y}$ monitor. This provides
the $S_{z}$ component of the first configuration. We verify that
$S_{x}$ is unaffected by the change of $\theta,B$, by that confirming
that the parameter change is appropriate.

\subsection*{Spin dynamics with polarization-modulated light: }

For small modulation depths $\theta\ll1$, the dynamics is governed
by Bloch-like equations of the vector $\left\langle \mathbf{F}\right\rangle =(F_{x},F_{y},F_{z})$
(see Supplementary Note for the general treatment). The spin orientations
$F_{x}$ (along the optical axis) and $F_{y}$ are subject to

\begin{eqnarray}
\dot{F}_{x} & = & -\gamma_{\perp}F_{x}-\omega_{\mathrm{B}}F_{y}-\Gamma\sin\left(2\theta\right)\cos\left(\omega t\right),\label{eq:Fx dynamics}\\
\dot{F}_{y} & = & -\gamma_{\perp}F_{y}+\omega_{\mathrm{B}}F_{x}+\Gamma\sin\left(2\theta\right)\sin\left(\omega t\right)F_{z},\label{eq:Fy dynamics}
\end{eqnarray}
which include a transverse decay rate $\gamma_{\perp}=\gamma+2\Gamma\cos^{2}\theta$
and Larmor precession at the rate $\omega_{\mathrm{B}}$. Here $\gamma$
denotes a slow ground-state depolarization rate (\textit{e.g.}, due
to wall collisions). The third term in Eq.~(\ref{eq:Fx dynamics})
is due to $s_{3}$. It describes a temporally-modulated optical pumping,
which is maximal at $\omega t=0$ (and at all $\omega t=2\pi n$ for
any integer $n$) towards $-\hat{x}$ (Fig.~\ref{fig: Scheme_spheres}a)
and at $\omega t=\pi$ towards $+\hat{x}$ (Fig.~\ref{fig: Scheme_spheres}c).
The pumping of $F_{x}$ is thus most efficient when the optical modulation
is synchronous with the Larmor precession $\omega=\omega_{\mathrm{B}}$.
The third term in Eq.~(\ref{eq:Fy dynamics}) is due to the modulated
linear polarization component $s_{2}$. It describes a tensor light-shift,
which acts as a magnetic field along $\hat{x}$ that rotates the spins
in the $yz$ plane at a modulated rate $\Gamma\sin\left(2\theta\right)\sin\left(\omega t\right)$.
The orientation $F_{z}$ along the magnetic field, which we aim to
generate, is subject to

\begin{equation}
\dot{F}_{z}=-\gamma_{\parallel}F_{z}-\Gamma\sin\left(2\theta\right)\left[\sin\left(\omega t\right)F_{y}-\cos\left(\omega t\right)\left\{ F_{z},F_{x}\right\} \right],\label{eq:Fz dynamics}
\end{equation}
where the first term is a longitudinal decay at a rate $\gamma_{\parallel}=\gamma+2\Gamma\sin^{2}\theta$,
the second term is again light shift due to $s_{2}$, and the third
term is an alignment-induced shift. The temporal modulation $\sin\left(\omega t\right)$
of the light shift is a key ingredient in pumping $\mathbf{F}$ towards
$+\hat{z}$, as it breaks the symmetry between the $\pm\hat{z}$ directions:
The sign of the light shift changes together with the sign of $F_{y}$,
thus acting as an alternating magnetic field that always tilts the
spins towards $+\hat{z}$, with maximal tilting rate obtained at $\omega t=\pi/2$
(Fig.~\ref{fig: Scheme_spheres}b) and $\omega t=3\pi/2$ (Fig.~\ref{fig: Scheme_spheres}d).
The tensor term $\{F_{z},F_{x}\}\equiv\bigl\langle\left\{ \mathbf{F}\hat{z},\mathbf{F}\hat{x}\right\} \bigr\rangle$
contributes similar spin buildup in amplitude but with a $\pi/2$
delay (see Supplementary Note). For $\omega=\omega_{\mathrm{B}}$,
both the synchronous pumping and the light shift are most efficient,
driving the precessing spins away from the $xy$ plane, transversely
to the optical axis. 

\onecolumngrid \appendix 
\setcounter{equation}{0}
\setcounter{figure}{0}
\setcounter{table}{0}
\setcounter{page}{1}
\makeatletter
\renewcommand{\theequation}{S\arabic{equation}}
\renewcommand{\thefigure}{S\arabic{figure}}
\renewcommand{\bibnumfmt}[1]{[S#1]}
\renewcommand{\citenumfont}[1]{S#1} 

\part*{Supplementary Information }

In this supplementary material, we derive the Bloch equations {[}Eqs.~(2-4)
in the main text{]} of $I=1/2$ spins interacting with polarization-modulated
light. The dynamics of the atoms can be described by the open quantum
system Liouville equation

\begin{equation}
\frac{d\rho}{dt}=-i\left[\mathcal{H}_{0},\rho\right]+\underset{\text{SE}}{\underbrace{\gamma_{\mathrm{e}}\sum_{i=-1}^{1}\mathcal{L}\left(\left|i\right\rangle \left\langle e\right|,\rho\right)}}+\underset{\text{SD}}{\underbrace{\gamma\sum_{i=-1}^{1}\mathcal{L}\left(S_{i},\rho\right)}},\label{eq: Lindblad dynamics-1}
\end{equation}
where $\rho$ is the density matrix of a single spin, and $\mathcal{L}$
is the Lindblad super-operator given by
\[
\mathcal{L}\left(A,\rho\right)=A\rho A^{\dagger}-\left(A^{\dagger}A\rho+\rho A^{\dagger}A\right)/2.
\]
The first term in Eq.~(\ref{eq: Lindblad dynamics-1}) denotes the
free Hamiltonian evolution; The second term denotes the excited state
decay due to spontaneous emission (SE) at a rate $\gamma_{\mathrm{e}}$;
and the last term denotes the ground-state relaxation due to spin
destruction (SD) at a rate $\gamma$, caused by interaction of the
electronic spin operator $\boldsymbol{S}$ with some thermal bath.
The free Hamiltonian term $\mathcal{H}_{0}$ describes the interaction
with external magnetic fields and laser beams. It is given by

\begin{align}
\mathcal{H}_{0} & =\omega_{0}\left|e\right\rangle \left\langle e\right|+\omega_{\mathrm{B}}\left(\left|1\right\rangle \left\langle 1\right|-\left|-1\right\rangle \left\langle -1\right|\right)\label{eq: Bare Hamiltonian}\\
+ & \Omega\bigl[\cos\left(\theta\right)e^{i\omega_{L}t}\left|0\right\rangle \left\langle e\right|+(1/\sqrt{2})\sin\left(\theta\right)e^{i\left(\omega_{L}+\omega\right)t}\left(\left|-1\right\rangle \left\langle e\right|+\left|1\right\rangle \left\langle e\right|\right)+\text{H.c.}\bigr],\nonumber 
\end{align}
where $\omega_{0}$ is the optical transition to the excited state,
$\omega_{\mathrm{B}}=gB/(2\hbar)$ is the Larmor frequency induced
by the magnetic field $B\hat{z}$, $\omega_{L}$ is the laser oscillation
rate, $\omega$ is the polarization-modulation rate and $\sin\left(\theta\right)$
is the modulation depth. $\Omega=E_{0}d/\hbar$ is pump laser Rabi
frequency where  $\mathbf{d}$ is the atomic dipole-moment and $E_{0}$
is the electric field amplitude of the pump beam. The matrix super-operator
representation of the SE term in Eq.~(\ref{eq: Lindblad dynamics-1})
is given by 

\begin{equation}
\left(\frac{d\rho}{dt}\right)_{\text{SE}}=\gamma_{\mathrm{e}}\left(\begin{array}{cccc}
\frac{1}{3}\rho_{ee} & 0 & 0 & -\frac{1}{2}\rho_{1e}\\
0 & \frac{1}{3}\rho_{ee} & 0 & -\frac{1}{2}\rho_{0e}\\
0 & 0 & \frac{1}{3}\rho_{ee} & -\frac{1}{2}\rho_{-1e}\\
-\frac{1}{2}\rho_{e1} & -\frac{1}{2}\rho_{e0} & -\frac{1}{2}\rho_{e-1} & -\rho_{ee}
\end{array}\right)\label{eq:spontaneous-emission}
\end{equation}
where the matrix elements of $\rho$ are ordered with the basis $\left\{ \left|1\right\rangle ,\left|0\right\rangle ,\left|-1\right\rangle ,\left|e\right\rangle \right\} \otimes\left\{ \left\langle 1\right|,\left\langle 0\right|,\left\langle -1\right|,\left\langle e\right|\right\} $.
Finally, the matrix representation of the SD term in Eq.~(\ref{eq: Lindblad dynamics-1})
is given by

\begin{equation}
\left(\frac{d\rho}{dt}\right)_{\text{SD}}=\gamma\left(\begin{array}{cccc}
\left(-\rho_{11}+\frac{1}{2}\rho_{00}+\frac{1}{8}\right) & -\frac{3}{2}\rho_{10}+\frac{1}{2}\rho_{0-1} & -2\rho_{1-1} & 0\\
-\frac{3}{2}\rho_{01}+\frac{1}{2}\rho_{-10} & \left(-\frac{3}{2}\rho_{00}+\frac{1}{2}\left(\rho_{11}+\rho_{-1-1}\right)+\frac{1}{4}\right) & -\frac{3}{2}\rho_{0-1}+\frac{1}{2}\rho_{10} & 0\\
-2\rho_{-11} & -\frac{3}{2}\rho_{-10}+\frac{1}{2}\rho_{01} & \left(-\rho_{-1-1}+\frac{1}{2}\rho_{00}+\frac{1}{8}\right) & 0\\
0 & 0 & 0 & 0
\end{array}\right).\label{eq:S-damping}
\end{equation}
In the derivation of this term, we assumed that the lower hyperfine
level $F_{g}=0$ is not populated, due to the action of a strong repump
beam. 

We can simplify the solution by transforming to a reference frame
rotating at the laser frequency $\omega_{L}$. Assuming a resonant
optical transition $\omega_{L}=\omega_{0}$, the Hamiltonian in the
rotating frame is given by

\begin{equation}
\mathcal{H}_{0}\rightarrow\omega_{\mathrm{B}}\left(\left|1\right\rangle \left\langle 1\right|-\left|-1\right\rangle \left\langle -1\right|\right)+\Omega\left[\cos\left(\theta\right)\left|0\right\rangle \left\langle e\right|+(1/\sqrt{2})\sin\left(\theta\right)e^{i\omega t}\left(\left|-1\right\rangle \left\langle e\right|+\left|1\right\rangle \left\langle e\right|\right)+\text{\text{H.c.}}\right],\label{eq:rotated H0}
\end{equation}
whereas the Liouville terms in Eqs.~(\ref{eq:spontaneous-emission})-(\ref{eq:S-damping})
are invariant under the transformation. Much below saturation ($\Omega\ll\gamma_{\mathrm{e}}$),
the excited state population is small ($\rho_{ee}\ll1$), and we can
adiabatically eliminate the excited state. This elimination results
with the steady-state coherences

\begin{align}
\rho_{e1} & \approx-2i\frac{\Omega}{\gamma_{\mathrm{e}}}\left(\cos\left(\theta\right)\rho_{01}+\nicefrac{1}{\sqrt{2}}\sin\left(\theta\right)e^{-i\omega t}\left(\rho_{11}+\rho_{-11}-\rho_{ee}\right)\right)\label{eq:rho41}\\
\rho_{e0} & \approx-2i\frac{\Omega}{\gamma_{\mathrm{e}}}\left(\cos\left(\theta\right)\left(\rho_{00}-\rho_{ee}\right)+\nicefrac{1}{\sqrt{2}}\sin\left(\theta\right)e^{-i\omega t}\left(\rho_{10}+\rho_{-10}\right)\right)\label{eq:rho42}\\
\rho_{e-1} & \approx-2i\frac{\Omega}{\gamma_{\mathrm{e}}}\left(\cos\left(\theta\right)\rho_{0-1}+\nicefrac{1}{\sqrt{2}}\sin\left(\theta\right)e^{-i\omega t}\left(\rho_{-1-1}+\rho_{1-1}-\rho_{ee}\right)\right)\label{eq:rho43}
\end{align}
and steady-state population 
\begin{equation}
\rho_{ee}\approx i\frac{\Omega}{\gamma_{\mathrm{e}}}\left(\cos\left(\theta\right)\rho_{e0}+\nicefrac{1}{\sqrt{2}}\sin\left(\theta\right)e^{i\omega t}\left(\rho_{e1}+\rho_{e-1}\right)+\text{c.c.}\right).\label{eq:rho44}
\end{equation}
We now use Eqs.~(\ref{eq: Lindblad dynamics-1}), (\ref{eq:spontaneous-emission}-\ref{eq:rotated H0}),
and (\ref{eq:rho41})-(\ref{eq:rho44}) to derive the Bloch equations.
The alignment of the spins is determined from 

\begin{align}
\frac{d}{dt}F_{z}^{2} & =\dot{\rho}_{11}+\dot{\rho}_{-1-1}=\left(\frac{8}{3}\Gamma\cos^{2}\left(\theta\right)+\frac{3}{2}\gamma\right)-\left(2\gamma+\frac{2}{3}\frac{\Omega^{2}}{\gamma_{\mathrm{e}}}\left(4\cos^{2}\left(\theta\right)+\sin^{2}\left(\theta\right)\right)\right)F_{z}^{2}\\
+ & \frac{\Gamma}{3}\sin\left(2\theta\right)\left(i\left[F_{z},F_{y}\right]\mbox{cos}\left(\omega t\right)-\left\{ F_{z},F_{y}\right\} \mbox{sin}\left(\omega t\right)\right),\nonumber 
\end{align}
where $\Gamma\equiv\Omega^{2}/\gamma_{\mathrm{e}}$ is the optical
pumping rate. For small modulation depths $\theta\ll1$, the last
term is negligible and we can describe the alignment build up by 
\[
\dot{F}_{z}^{2}=-\left(R_{a}+\gamma/2\right)F_{z}^{2}+R_{a},
\]
where $R_{a}=\frac{8}{3}\Gamma\cos^{2}\left(\theta\right)+\frac{3}{2}\gamma$.
At steady state $F_{z}^{2}\approx1$ for $R_{a}\gg\gamma.$ The oscillating
spin orientation $F_{+}=\frac{1}{\sqrt{2}}\left(F_{x}+iF_{y}\right)$
obeys equation

\begin{align}
\dot{F}_{+} & =\dot{\rho}_{01}+\dot{\rho}_{-10}=\left(i\omega_{\mathrm{B}}-\gamma-\Gamma\left(2\cos^{2}\left(\theta\right)+\sin^{2}\left(\theta\right)\right)\right)F_{+}\label{eq:F_plus_equation}\\
- & \frac{\Gamma}{\sqrt{2}}\sin\left(2\theta\right)\left(\left(2-F_{z}^{2}\right)\mbox{cos}\left(\omega t\right)-iF_{z}\mbox{sin}\left(\omega t\right)\right)-\Gamma\sin^{2}\left(\theta\right)F_{-},\nonumber 
\end{align}
where $F_{-}=\frac{1}{\sqrt{2}}\left(F_{x}-iF_{y}\right)$. Again
for small modulation depths $\theta\ll1$, the last term, proportional
to $\sin^{2}\left(\theta\right)$, is negligible. By separating the
real and imaginary parts of Eq.~(\ref{eq:F_plus_equation}), we derive
the equations of motion of $F_{x}$ and $F_{y}$ given in the main
text

\begin{equation}
\dot{F}_{x}=-\left(\gamma+\Gamma\left(2\cos^{2}\left(\theta\right)+\sin^{2}\left(\theta\right)\right)\right)F_{x}-\omega_{\mathrm{B}}F_{y}-\Gamma\sin\left(2\theta\right)\cos\left(\omega t\right)\left(2-F_{z}^{2}\right),\label{eq:Fx dynamics-1}
\end{equation}

\begin{equation}
\dot{F}_{y}=-\left(\gamma+\Gamma\left(2\cos^{2}\left(\theta\right)+\sin^{2}\left(\theta\right)\right)\right)F_{y}+\omega_{\mathrm{B}}F_{x}+\Gamma\sin\left(2\theta\right)\sin\left(\omega t\right)F_{z}.\label{eq:Fy dynamics-1}
\end{equation}
Here we can identify $\gamma_{\perp}=\gamma+2\Gamma\cos^{2}\theta$
as the effective 'transverse' destruction rate, used in the main text,
and we consider the steady state $F_{z}^{2}\approx1.$ Finally, the
dynamics of the spin orientation along the magnetic field $F_{z}$
is given by

\begin{equation}
\dot{F}_{z}=\dot{\rho}_{11}-\dot{\rho}_{-1-1}=-\left(2\Gamma\sin^{2}\left(\theta\right)+\gamma\right)F_{z}-\Gamma\sin\left(2\theta\right)\left(\mbox{cos}\left(\omega t\right)\left\{ F_{z},F_{x}\right\} -i\mbox{sin}\left(\omega t\right)\left[F_{z},F_{x}\right]\right).\label{eq:Fz dynamics-1}
\end{equation}
This equation corresponds to that used in the main text, with $\gamma_{\parallel}=\gamma+2\Gamma\sin^{2}\theta$
the effective 'longitudinal' destruction rate.

The mean alignment term $\left\{ F_{z},F_{x}\right\} $ in Eq.~(\ref{eq:Fz dynamics-1})
makes the above set of equations incomplete. This term is governed
by two additional coupled equations

\begin{align}
\frac{d}{dt}\left(F_{z}F_{x}+F_{x}F_{z}\right) & =-\frac{1}{\sqrt{2}}\left(\dot{\rho}_{-10}-\dot{\rho}_{01}+\dot{\rho}_{0-1}-\dot{\rho}_{10}\right)\label{eq:FzFx}\\
= & -2\left(\gamma+\Gamma\right)\left(F_{z}F_{x}+F_{x}F_{z}\right)-\omega_{\mathrm{B}}\left(F_{z}F_{y}+F_{y}F_{z}\right)-\Gamma\sin\left(2\theta\right)\mbox{cos}\left(\omega t\right)F_{z},\nonumber 
\end{align}
and
\begin{align}
\frac{d}{dt}\left(F_{z}F_{y}+F_{y}F_{z}\right) & =\frac{i}{\sqrt{2}}\left(\dot{\rho}_{-10}-\dot{\rho}_{0-1}-\dot{\rho}_{01}+\dot{\rho}_{10}\right)\label{FzFy}\\
= & -2\left(\gamma+\Gamma\right)\left(F_{z}F_{y}+F_{y}F_{z}\right)+\omega_{\mathrm{B}}\left(F_{z}F_{x}+F_{x}F_{z}\right)+\Gamma\sin\left(2\theta\right)\mbox{sin}\left(\omega t\right)\left(2-F_{z}^{2}\right).\nonumber 
\end{align}
Equations (\ref{eq:FzFx}) and (\ref{FzFy}) are mathematically similar
to Eqs.~(\ref{eq:Fy dynamics-1}) and (\ref{eq:Fx dynamics-1}),
respectively. We therefore conclude that the contribution of the alignment
term in Eq.~(\ref{eq:Fz dynamics-1}) is similar to that of the tensor
light-shift in Eqs.~(\ref{eq:Fy dynamics-1}) and (\ref{eq:Fx dynamics-1}),
only with a phase lag of $\pi/2$. 

\begin{thebibliography}{10}
\bibitem{footnote1}Either at very high magnetic fields ($\gtrsim$Tesla),
where the magnetic levels are optically resolved \citep{Walker-high-magnetic-fields},
or at zero magnetic field during pumping, followed by magnetic field
pulses for spin rotation.
\bibitem{Footnote3}This interaction is mediated via absorption and
stimulated emission of two co-propagating photons, avoiding spontaneous
emission.
\end{thebibliography}
\end{document}